# Lung Ultrasound Segmentation and Adaptation between COVID-19 and Community-Acquired Pneumonia

Harry Mason[1,2], Lorenzo Cristoni[3], Andrew Walden[4], Roberto Lazzari[5], Thomas Pulimood[6,7], Louis Grandjean[8,9], Claudia AM Gandini Wheeler-Kingshott[10,11], Yipeng Hu[1,2], Zachary MC Baum[1,2]

[1] Centre for Medical Image Computing, University College London
[2] Wellcome/EPSRC Centre for Surgical & Interventional Sciences, University College London
[3] Frimley Park Hospital, Frimley Health NHS Foundation Trust
[4] Royal Berkshire Hospital, Royal Berkshire NHS Foundation Trust
[5] Hospital de La Santa Creu I Sant Pau, Barcelona
[6] West Suffolk Hospital, West Suffolk NHS Foundation Trust
[7] Cambridge University Hospital, University of Cambridge
[8] Great Ormond Street Children's Hospital NHS Foundation Trust
[9] Institute of Child Health, University College London
[10] NMR Research Unit, Queen Square MS Centre, UCL Queen Square Institute of Neurology
[11] Department of Brain and Behavioural Sciences, University of Pavia
`harry.mason.18@ucl.ac.uk`

**Abstract.** Lung ultrasound imaging has been shown effective in detecting typical patterns for interstitial pneumonia, as a point-of-care tool for both patients with COVID-19 and other community-acquired pneumonia (CAP). In this work, we focus on the hyperechoic B-line segmentation task. Using deep neural networks, we automatically outline the regions that are indicative of pathology-sensitive artifacts and their associated sonographic patterns. With a real-world data-scarce scenario, we investigate approaches to utilize both COVID-19 and CAP lung ultrasound data to train the networks; comparing fine-tuning and unsupervised domain adaptation. Segmenting either type of lung condition at inference may support a range of clinical applications during evolving epidemic stages, but also demonstrates value in resource-constrained clinical scenarios. Adapting real clinical data acquired from COVID-19 patients to those from CAP patients significantly improved Dice scores from 0.60 to 0.87 ($p < 0.001$) and from 0.43 to 0.71 ($p < 0.001$), on independent COVID-19 and CAP test cases, respectively. It is of practical value that the improvement was demonstrated with only a small amount of data in both training and adaptation data sets, a common constraint for deploying machine learning models in clinical practice. Interestingly, we also report that the inverse adaptation, from labelled CAP data to unlabeled COVID-19 data, did not demonstrate an improvement when tested on either condition. Furthermore, we offer a possible explanation that correlates the segmentation performance to label consistency and data domain diversity in this point-of-care lung ultrasound application.

# 1 Introduction

Over the past decade, the use of point of care ultrasound (POCUS) has increased alongside the growing evidence relating its use to improved patient outcomes. The publication of the BLUE protocol displayed the efficiency of POCUS in the diagnosis of the 5 most common lung pathologies compared to chest auscultation and chest x-ray, achieving an accuracy of 90.5% [1]. POCUS was shown to be useful in the triaging of patients with suspected COVID-19 by following the BLUE protocol [2, 3]. Both COVID-19 and CAP present multiple B-lines in the early stages and areas of consolidation appear as infection progresses. Although computerized tomography (CT) scans have shown sensitivity of up to 97% [4] for the diagnosis of COVID-19, it can be impractical for use in 'front-line' settings, as it requires patients to be moved throughout the hospital, may risk precautious patients desaturating in scanner, and is time-consuming. Conversely, the BLUE protocol can be performed in a few minutes at the patient's bedside, making POCUS advantageous for use during a pandemic when resources are low and infection risk is high.

Several studies have investigated the use of deep learning to assist in triage, diagnosis, grading and monitoring of COVID-19 patients [5-10]. Methods include classifying and stratifying COVID-19 patients, or localizing pathological image features, all based on lung ultrasound (LUS) data from healthy subjects or other respiratory diseases, such as pulmonary edema and community-acquired pneumonia (CAP). To improve the specificity of computer-assisted tools, aggregating approaches combining pixel-, frame-, zone-, and patient-level severity scores have been proposed [11]. Localization, and therefore, segmentation, of pathology-sensitive LUS features then, on the pixel level, is fundamental. Moreover, the intuitive representation of segmentation, such as those of B-lines used in this study, may provide a visually interpretable solution in the form of a prediction for the clinician. Such segmentations may not only help the confidence the clinician would place on the automated computer prediction by localizing it, but also provides a feedback opportunity to further develop the assistive algorithm for improved sensitivity and specificity.

Most existing research in machine learning, such as the work we present here, requires retrospectively labelled data. However, deploying such algorithms in real-world clinical use has direct challenges. Most prominently, efficiently obtaining high-quality labelled data [12]. For example, at the beginning of an epidemic, or during its fast-changing stage, representative imaging data from positive patients is usually scarce. Furthermore, obtaining expert labels may be even more costly during the peak of an outbreak. In scenarios such as these, the ability to use a pre-trained model, or an existing data set, perhaps from a relevant condition (CAP, in this work), could substantially reduce the requirements for necessary data and labeling from the target application, using fine-tuning or unsupervised domain adaptation. A different type of scenario, also investigated in this work, may be that data from a previous or ongoing epidemic (COVID-19, in this work) are available for training the pre-trained models or being used as the source domain data to be adapted to a different type of condition that has less, limited, or unlabeled data. Examples include pneumonia caused by a new epidemic, an additional variant to the existing one, or other types of pneumonia in an area that lacks access to other data sources or labeling expertise. In this study, we test the transfer learning and domain adaptation abilities to and from the COVID-19 patient data, with the CAP patient data as an example of the other LUS data.

## 2      Methods

We consider two strategies for training convolutional neural networks to segment B-lines from LUS images. The first strategy uses a supervised learning approach, requiring manual labeling of all input data, to segment the B-lines in the LUS images. The second strategy uses an unsupervised domain adaptation to adapt a segmentation network to an unlabeled target domain, requiring labels for only the source segmentation domain in training.

### 2.1      Supervised Segmentation with U-Net

A commonly used neural network for image segmentation, U-Net [13], was trained to automatically segment B-lines in COVID-19 and CAP LUS images. At inference, the network then predicts whether a given pixel in the image may be classified as part of a B-line. The use of well-established network architectures, such as U-Net, allows this work to focus primarily on investigating the feasibility of automatic segmentation of these regions of pathological interest.

### 2.2      Unsupervised Segmentation via Image and Feature Alignment

Synergistic image and feature alignment (SIFA) [14] has been used for domain adaptation tasks to guide the adversarial learning of an end-to-end framework for unsupervised image segmentation. SIFA reduces domain shift by using a generative adversarial network to synthetically translate images from a source domain to the target domain. The network is composed of a generator, which learns to translate the source domain image into a corresponding image of the target domain, an encoder that learns a shared feature-space, a decoder that learns the reverse-translation from target to source, and a segmenter that performs pixel-wise classification to identify different labels in the images. Additionally, three discriminator networks differentiate between the target and source inputs to the encoder, and the outputs of the decoder and segmenter. SIFA is trained to automatically segment B-lines in COVID-19 and CAP LUS images. However, in training, labels for only the source domain are required to learn the segmentation of the target domain.

### 2.3      Implementation Details

All neural networks were implemented in TensorFlow [15] and Keras [16]. Reference-quality open-source code was adopted where possible for reproducibility.

Our implementation of U-Net contained 4 layers of convolutional blocks, with an increasing number of channels of 16, 32, 64, and 128. Each convolutional block used

Batch Normalization across the channel axis between convolutional layers and a Dropout of 0.5 following each Batch Normalization. We employed data augmentation using rotation, shifting, and scaling to reduce over-fitting. All U-Net models were trained for 250 epochs with a mini-batch size of 16, using an equal-weight binary cross-entropy and Dice loss and the Adam optimizer [17] with a learning rate of 0.005.

Our implementation of SIFA and hyperparameters described below are consistent with the original default implementation and hyperparameters, as described in [14]. As in the original implementation, we employed data augmentation using rotation, shifting, and scaling to reduce over-fitting. All SIFA models were trained for 10,000 epochs, with a mini-batch size of 12. The generator, encoder, and decoder were trained using a weighted cycle-consistency and adversarial loss with the Adam optimizer at a learning rate of 0.0002. The segmenter was trained using an equal-weight cross-entropy and Dice loss with the Adam optimizer at a learning rate of 0.001.

### 2.4 Data

The US images were acquired from two hospitals by two clinicians, using a Butterfly iQ US probe (Butterfly Inc., Guilford, CT, USA). Experiments were conducted using images from six COVID-19 positive patients and seven patients with CAP. Due to the low prevalence of B-lines within patient scans only images with B-lines were used for training, to evaluate the segmentation algorithms. The resulting datasets contained 977 and 326 images for COVID-19 and CAP, respectively. All COVID-19 diagnoses were confirmed by PCR tests.

Ground-truth B-line segmentations were manually labeled by a medical student familiar with LUS. Segmentations were reviewed and verified by experienced US imaging researchers with over five years of experience with clinical US imaging. Example images and segmentations are provided in Figure 1.

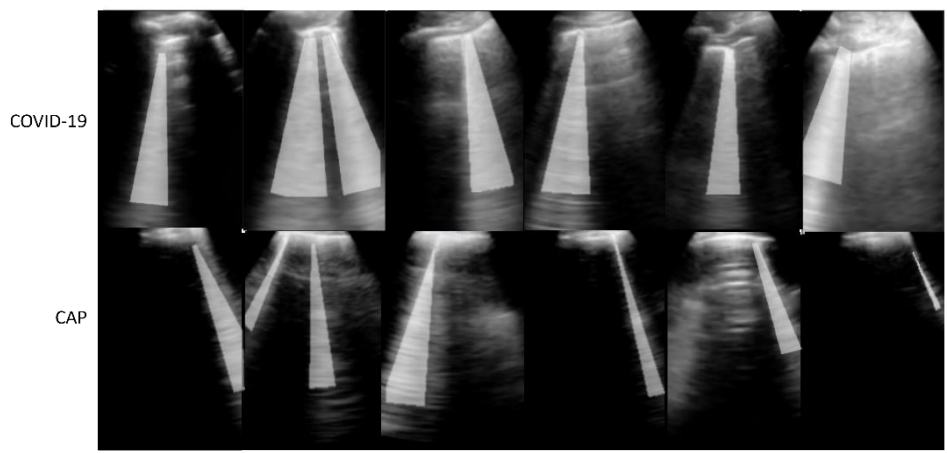

**Fig. 1.** Multiple sample LUS images and their corresponding manual segmentations. COVID-19 images and segmentations are shown on the top row, CAP images and segmentations are shown on the bottom row.

**2.5 Experiments**

Given the limited data availability, we adopt a two-way split for training and test sets, without a validation set. This prevents fine-tuning of network parameters or other hyperparameters to optimize performance, to avoid information leakage and provide fair estimates of model performances. As such, cross-validation was performed to assess the performance of models for the segmentation of B-lines in patients with COVID-19 and CAP under different supervision conditions in training. The COVID-19 and CAP datasets were split into three cross-validation sets, on a patient-level. Splitting the data in this way ensures that no patient images are found amongst the different dataset splits. Efforts were made to ensure that each of the three COVID-19 and CAP datasets were of approximately the same size. The COVID-19 datasets each consisted of 2 patients, with 319, 319, and 339 images, respectively. The CAP datasets consisted of one, two, and three patient(s), with 148, 111, and 67 images, respectively. In total, seven experiments are presented to evaluate and assess the performance of U-Net and SIFA for segmentation of B-lines in patients with COVID-19 and CAP.

Four of these seven experiments are performed with U-Net. First, we train U-Net with COVID-19 images. Second, we train U-Net with CAP images. Third, we train U-Net with COVID-19 images and fine-tune with CAP images. Finally, we train U-Net with CAP images and fine-tune with COVID-19 images. In both instances, fine-tuning took place over 50 epochs at a learning rate of 0.0005. Corresponding COVID-19 and CAP datasets are used in training and fine-tuning when applicable.

Three of these seven experiments are performed with SIFA. First, we train SIFA with a source domain of COVID-19 images and a target domain of CAP images. To assess if the discrepancy in dataset sizes affects the training of SIFA, we then train SIFA with a source domain of COVID-19 images, where we use only a reduced subset of the COVID-19 images and a target domain of CAP images. Finally, we train SIFA with a source domain of CAP images and a target domain of COVID-19 images. To evaluate each of the previously described methods, segmentations were evaluated based on a binary Dice score. We additionally report sensitivity, specificity and p-values from statistical t-tests at a significance level of 0.05, when comparison was made.

**3  Results**

Table 1 summarizes the Dice scores from the cross-validation experiments across all methods. Training with SIFA (COVID-19 Source / CAP Target) provided significantly higher Dice scores on COVID-19 and CAP test data than all four U-Net methods ($p < 0.001$) and with SIFA (CAP Source / COVID-19 Target) ($p < 0.001$). Training with SIFA (CAP Source / COVID-19 Target) provided significantly lower Dice scores on the COVID-19 and CAP test data than U-Net (COVID-19) ($p < 0.001$), U-Net (CAP) ($p < 0.001$), and U-Net (COVID-19 Fine-Tune w/ CAP), ($p < 0.001$). Additionally, training with SIFA (CAP Source / COVID-19 Target) provided significantly lower Dice scores on the COVID-19 test data than U-Net (CAP Fine-Tune w/ COVID-19), ($p < 0.001$), but no significant difference was found to U-Net (CAP Fine-Tune w/ COVID-19) when applied to CAP test data ($p = 0.23$). Table 2 summarizes the sensitivity and specificity from the cross-validation experiments and is consistent with the observation summarized above.

**Table 1.** Summary of segmentation cross-validation Dice scores. STD: Standard Deviation. Values are presented as Mean ± STD.

| Method | COVID-19 Dice | CAP Dice |
|---|---|---|
| U-Net (COVID-19) | 0.60 ± 0.26 | 0.36 ± 0.24 |
| U-Net (CAP Fine-Tune w/ COVID-19) | 0.56 ± 0.25 | 0.35 ± 0.26 |
| U-Net (CAP) | 0.52 ± 0.17 | 0.43 ± 0.27 |
| U-Net (COVID-19 Fine-Tune w/ CAP) | 0.55 ± 0.15 | 0.45 ± 0.24 |
| SIFA (COVID-19 Source / CAP Target) | 0.87 ± 0.13 | 0.71 ± 0.22 |
| SIFA (Reduced COVID-19 Source / CAP Target) | 0.83 ± 0.15 | 0.72 ± 0.21 |
| SIFA (CAP Source / COVID-19 Target) | 0.32 ± 0.21 | 0.33 ± 0.17 |

**Table 2.** Summary of segmentation cross-validation sensitivity (sens.) and specificity (spec.).

| Method | COVID-19 Sens. | COVID-19 Spec. | CAP Sens. | CAP Spec. |
|---|---|---|---|---|
| U-Net (COVID-19) | 0.55 | 0.93 | 0.38 | 0.92 |
| U-Net (CAP Fine-Tune w/ COVID-19) | 0.47 | 0.97 | 0.44 | 0.95 |
| U-Net (CAP) | 0.50 | 0.90 | 0.50 | 0.93 |
| U-Net (COVID-19 Fine-Tune w/ CAP) | 0.48 | 0.91 | 0.40 | 0.95 |
| SIFA (COVID-19 Source / CAP Target) | 0.86 | 0.98 | 0.78 | 0.96 |
| SIFA (Reduced COVID-19 Source / CAP Target) | 0.84 | 0.95 | 0.75 | 0.95 |
| SIFA (CAP Source / COVID-19 Target) | 0.30 | 0.95 | 0.31 | 0.94 |

Additionally, we present qualitative examples of segmentations from both methods, trained with all the different aforementioned approaches, on COVID-19 images and CAP images in Figures 2 and 3, respectively. These visualizations demonstrate the ability of these networks to delineate B-lines, suggesting that, in some instances, they may be effectively used for assisting in the interpretation of LUS in clinical practice.

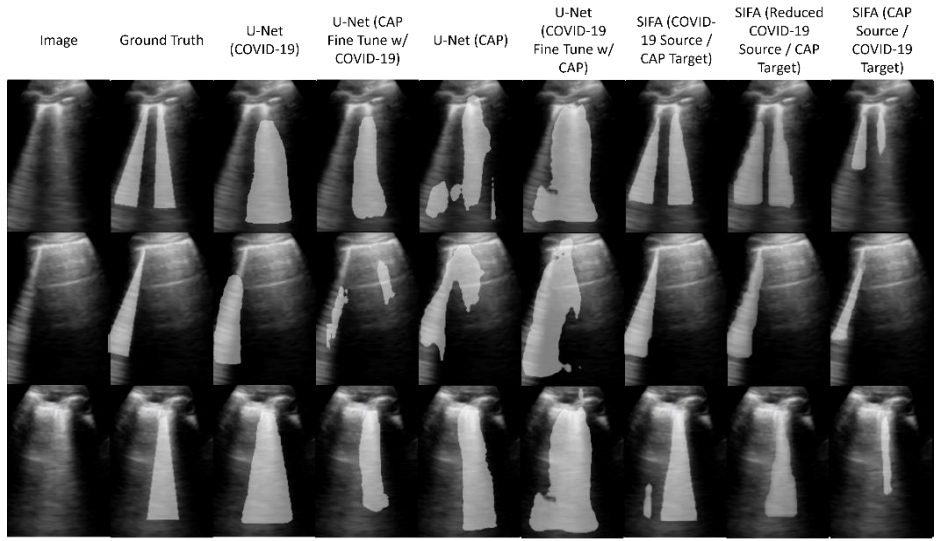

**Fig. 2.** Three example LUS images, each illustrating segmentations from each of the different methods on COVID-19 images. Each image shows the original LUS image and the segmentation output corresponding to the ground truth, or the method used. Each column presents the image and segmentation for the method listed above them.

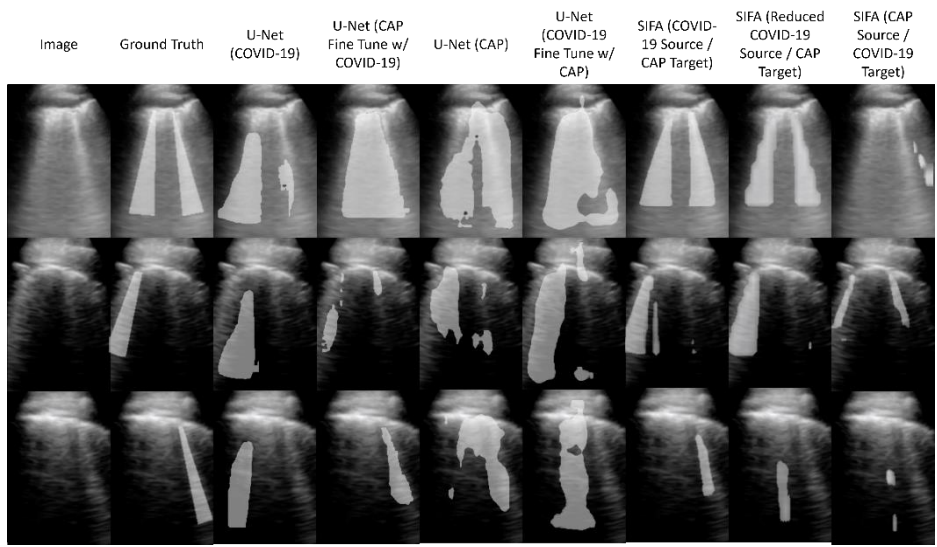

**Fig. 3.** Three example LUS images, each illustrating segmentations from each of the different methods on CAP images. Each image shows the original LUS image and the segmentation output corresponding to the ground truth, or the method used. Each column presents the image and segmentation for the method listed above them.

As a retrospective analysis, we aim to explain the observed difference in improvement (or lack of it) between the two directions of adaptation when testing the resulting models on both data sets, as described above, in terms of the difference in the imaging data and labels available to training and testing. Figure 1 provided examples images with their ground-truth segmentations overlaid, from the COVID-19 and CAP data sets, in the upper and lower rows, respectively. It is visibly evident that CAP data posses substantially higher variability in location, size of the identified B-line patterns and their background context. This is consistent with all the data used in our study. This is also consistent with the annotators' experience indicating that labelling on the CAP data set is considered a more challenging task than that on the COVID-19 data set.

## 4   Discussion

Additionally, during initial experimentation, we evaluated the performance of a joint-training strategy for supervised segmentation with U-Net in addition to the fine-tuning methods. Here, instead of fine-tuning on a pre-trained model, we train the model on both the COVID-19 and CAP datasets simultaneously. Notably, the performance over all cross-validation folds resulted in comparable Dice scores to training only on COVID-19 when tested on both COVID-19 and CAP test sets. For brevity, we did not include a full validation of this training strategy in our above-presented experiments.

One of the interesting findings in this work is that the substantial difference between the adaptation from two opposite directions, from CAP to COVID-19 and from COVID-19 to CAP, with only the latter showing benefit of adaptation on both test datasets. It is not unsurprising that the adapted models may outperform the models trained solely with individual datasets in a supervised manner. This is compounded by the fact that with this adaptation, there is additional data and data diversity. However, this is not consistent with the performance observed when adapting from CAP to COVID-19. Intuitively, we may associate this with the label uncertainty and variability observed within the CAP images, as previously discussed in Section 3. It is important to note that, especially constrained by small data sets, the efficacy of domain adaptation is highly dependent on the data diversity and label uncertainty, one needs to be further understood and validated before being deployed in clinical applications.

## 5   Conclusion

In this work, we have presented the development and validation experiments for segmenting real clinical LUS data, acquired from both CAP and COVID-19 patients, and in particular the approaches for combining the two for training deep neural networks. We report a set of interesting experimental results that demonstrated that, in a small data set setting, domain adaptation can be effective in improving segmentation accuracy by incorporating additional unlabelled data. However, compared to the direction of the desirable adapting, the availability of diverse data and high-quality, consistent and representative labels were more strongly correlated with such improvement. The experimental results provided preliminary evidence for the feasibility and practicality of aggregating different types of data in this POCUS application.

**Acknowledgments.** This work is supported by the Wellcome/EPSRC Centre for Interventional and Surgical Sciences (203145Z/16/Z). C.A.M. Gandini Wheeler-Kingshott is supported by the MS Society (#77), Wings for Life (#169111), Horizon2020 (CDS-QUAMRI, #634541), BRC (#BRC704/CAP/CGW), and allocation from the UCL QR Global Challenges Research Fund (GCRF). Z.M.C. Baum is supported by the Natural Sciences and Engineering Research Council of Canada Postgraduate Scholarships-Doctoral Program, and the University College London Overseas and Graduate Research Scholarships.